\documentstyle[12pt]{article}
\begin{document}
\setlength{\baselineskip}{0.30in}
\newcommand{\beq}{\begin{equation}}
\newcommand{\eeq}{\end{equation}}
\newcommand{\bi}{\bibitem}

{\hbox to\hsize{August, 1996  \hfill TAC-1996-021}
\begin{center}
\vglue .06in
{\Large \bf { Higher spin fields and the problem of
 cosmological constant.}
}
\bigskip
\\{\bf A.D. Dolgov}
 \\[.05in]
{{Teoretisk Astrofysik Center\\
 Juliane Maries Vej 30, DK-2100, Copenhagen, Denmark
\footnote{Also: ITEP, Bol. Cheremushkinskaya 25, Moscow 113259, Russia.}
}}\\[.40in]

\end{center}
\begin{abstract}

The cosmological evolution of free massless vector or tensor (but not gauge)
fields minimally coupled to gravity is analyzed. It is shown that there are
some unstable solutions for these fields in De Sitter background. The back
reaction of the energy-momentum tensor of such solutions to the original
cosmological constant exactly cancels the latter and the expansion regime
changes from the exponential to the power law one. In contrast to the
adjustment
mechanism realized by a scalar field the gravitational coupling constant in
this
model is time-independent and the resulting cosmology may resemble the
realistic one.

\end{abstract}
%\newpage
\section{Introduction}

The mystery of the cosmological constant $\Lambda$ is one of the most profound
or just the most profound one in modern fundamental physics. Astronomical
observations show that it is extremely small on the scale of elementary
particle physics while it still may be cosmologically essential. The
astronomical bounds (see e.g. the review\cite{cpt})
on the vacuum energy density
$\rho_{vac} = \Lambda m_{Pl}^2 /8\pi$
are roughly speaking that $\rho_{vac}$ does not exceed the value of the
critical energy density $\rho_c = 2 \cdot 10^{-29} h_{100}^2\> {\rm g/cm^3} $.
(Here $h_{100}$ is the Hubble parameter in units 100 km/sec/Mpc.) Written in
terms of particle physics units the bounds reads:
 \beq{
 \rho_{vac} < 10^{-47} {\rm GeV}^4
\label{rhovac}
}\eeq

On the other hand there are plenty contributions coming from different physical
fields which by many orders of magnitude exceed the permitted value
(\ref{rhovac}). For the review see papers \cite{sw1,ad1}. For example the
energy density of the chiral condensate $\langle \bar q q \rangle$
well established in Quantum Chromodynamics (QCD)\cite{gor} is approximately
$\rho_{qq} = 10^{-3} - 10^{-4}\> {\rm GeV}^4$ and the energy density of gluon
condensate\cite{svz} is $\rho_{GG} = 10^{-3} - 10^{-4}\> {\rm GeV}^4$
which are at least by 44-45 (!)
orders of magnitude larger than the astronomical bound on the vacuum
energy (\ref{rhovac}). There could be some other contributions which are even
much bigger than $\rho_{qq}$ or $\rho_{GG}$.
In particular supergravity or superstring models
naturally imply $\rho_{vac} \approx m_{Pl}^4/(8\pi)^2$ which exceeds the
bound (\ref{rhovac}) by approximately 120 orders of magnitude.

It is hard, or better to say, impossible to
believe to an accidental cancellation with such a precision by a contribution
from some other fields, which know nothing about quarks and gluons,
so one is forced
to find a mechanism which can somehow achieve that dynamically.
This is definitely the
problem of low energy physics because the characteristic scale at which this
mechanism should operate is above
$l= \rho_{vac}^{1/4} \approx (10^{12} {\rm GeV})^{-1} \approx 10^{-2}{\rm cm}$.
The natural idea is to invent an adjustment
mechanism \cite{ad2,wil,psw,lf,bh,fuj} realized by
a new massless (or extremely light) classical field with not necessarily
positive definite energy density. The interaction of this
field with the curvature of space-time should be chosen in such a way that its
energy-momentum tensor would cancel down the underlying vacuum energy. It
resembles the axionic solution of the problem of strong CP-violation in
QCD \cite{pq, sw2, fw}. However the attempts to realize the adjustment with
a scalar field proved to be unsuccessful. The free massless
scalar field obeying the equation of motion $D^2 \phi = 0$ (where $D$ is the
covariant derivative in the gravitational background) is stable, or in other
words it does not possess solution rising with time and its stress tensor
asymptotically vanishes. However a scalar field
non-minimally coupled to the curvature as $\xi R \phi^2$ is indeed
unstable in De Sitter space-time (if $\xi R <0$)\cite{ad2}
and its back reaction turns
the exponential expansion, $a(t) \sim \exp (Ht)$ into a power law one,
$a(t) \sim t^\sigma$ but at the expense of the asymptotically vanishing
gravitational constant $G_N \sim 1/t^2$. Moreover the energy-momentum tensor
of such scalar field is not proportional to the vacuum energy-momentum tensor
$T^{vac}_{\mu\nu} = \rho_{vac} g_{\mu\nu}$ but has a quite different tensor
structure.

It has been argued\cite{ad2,ad3,ad1}
that despite the absence of satisfactory models one still
can conclude that an adjustment mechanism generically leads to a non-complete
cancellation of the vacuum energy with the non-compensated remnant of the order
of $\delta \rho \sim m_{Pl}^2/t^2$. The non-compensated part of the
energy-momentum tensor
$\delta T_{\mu\nu} = T^{(vac)}_{\mu\nu} - T^{(\phi)}_{\mu\nu}$
is not necessarily of the vacuum-like form
but may have an arbitrary, possibly an exotic, relation between
the pressure $\delta p$ and the energy $\delta \rho$ densities. Under a
simplifying assumption that  $\delta T_{\mu\nu}$ have the vacuum form,
that is $\delta T_{\mu\nu}\sim g_{\mu\nu}$ the idea of the time varying
vacuum energy density was lately explored in several papers\cite{lamt}
for construction of realistic cosmologies with time dependent cosmological
"constant". Further development along these lines is inhibited by a lack of
a consistent Lagrangian model without fine-tuning.

At first glance a scalar field is the only candidate for the adjustment
mechanism because its spatially constant classical condensate does not destroy
the observed homogeneity and isotropy of the universe. However this is not
necessarily the case. For example the time component of a vector field can
play this role without violating isotropy and homogeneity\cite{ad3,ad4}. In
ref.\cite{ad3} a gauge vector field with the usual kinetic term
$F^2_{\mu\nu}$ was considered. Such field is stable in the De Sitter background
and to induce an instability the coupling to the curvature which breaks
gauge symmetry was introduced, $\xi R U (A_\mu^2)$. The model contains too much
arbitrariness, connected with the choice of the potential $U(A^2)$,
and gives rise to a time dependent gravitational constant
though the dependence can be much milder than in the scalar case, e.g.
$G_N$ may logarithmically depend on time.

A more interesting model is based on the gauge non-invariant Lagrangian of the
form\cite{ad4}:
\beq{
 {\cal L}_0 =  \eta_0 A_{\alpha;\beta}A^{\alpha;\beta}
\label{la}
}\eeq
without any potential terms.
The classical equation of motion for the time component $A_t$ in this case has
indeed an unstable solution and with the proper sign of the constant $\eta_0$
the energy-momentum tensor corresponding to this solution compensates the
vacuum one. The non-compensated terms die down as $1/t^2$. Unfortunately the
cosmology based on this model is not realistic because the scale factor rises
too fast, $a(t) \sim t$. We will consider this  model in some detail in the
next section.

The set of possible compensating fields which do not break the isotropy and
homogeneity is not exhausted by a scalar $\phi$ and vector $A_\alpha$. There
can be higher rank tensor fields like e.g. time components of symmetric tensor
$S_{\alpha\beta}$ ($S_{tt}$) or even $S_{\alpha\beta\gamma}$ ($S_{ttt}$),
and isotropic space components like $S_{ij}\sim \delta_{ij}$ or
space components of antisymmetric tensor $A_{ijk}$. With the simplest
Lagrangian
analogous to (\ref{la}) these fields (which we denote generically as
$V_{\alpha\beta...}$) satisfy the equation of motion
\beq{
D^2 V_{ \alpha\beta...} = 0
\label{d2v}
}\eeq
As we see in what follows these fields are also unstable and can change the
De Sitter expansion to the power law one, and with a  particular choice
of the Lagrangian, one can get $H=1/2t$ or
(with a slightly different Lagrangian)  $H=2/3t$ which are already close
to realistic cosmologies. Higher rank symmetric tensor fields have not yet
been considered but antisymmetric ones naturally appear in high dimensional
supergravity or superstring models. These fields however are supposed to be
gauge fields but in this case they are stable, so we reject gauge invariance
from the very beginning.

It may be not an innocent assumption to introduce massless non-gauge fields and
moreover with non-positive definite energy density. There may be even some
more serious restrictions on the theory. For example in the case of the
vector field $A_{\alpha}$ one has to impose the condition of vanishing of
spatial components of the  field, $A_i = 0$, otherwise these components,
which are also unstable would destroy the anisotropy of the universe. This is
in a drastic contrast to the normal additional condition
$V^\alpha_{\>;\alpha}=0$
which kills the scalar component. It may create a lot of problems for
quantization of such a theory. However our aim here is more modest, it is
just to find possible classical solutions of equations of motion
which follow from relatively simple Lagrangians and which would
kill the cosmological constant. With higher rank tensor fields this problem
seem to be easily solved but it may be very difficult (if possible) to make
really realistic cosmology.
The solution discussed here presents at least a counter example to
the "no-go" theorem for the adjustment mechanism,
proposed by S. Weinberg in his review\cite{sw1}.

\section{Vector field.}

Here and in what follows we assume that the background metric is the spatially
flat Robertson-Walker one:
\beq{
 ds^2 = dt^2 -a^2(t) d\vec r\>^2
\label{ds2}
}\eeq
The equations of motion (\ref{d2v}) for the vector field $A_\mu$ in this metric
have the form:
\beq{
(\partial^2_t -{1\over a^2} \partial^2_j +3H \partial_t -3H^2) A_t
+{2H \over a^2} \partial_j A_j = 0,
\label{d2tat}
}\eeq
\beq{
(\partial^2_t -{1\over a^2} \partial^2_j +H \partial_t -\dot H - 3H^2) A_j
+2H \partial_j A_t = 0
\label{dt2aj}
}\eeq
where $H=\dot a/a$ is the Hubble parameter.

The energy-momentum tensor of this field is easily calculated from the
Lagrangian (\ref{la}) and is equal to:
\begin{eqnarray}
\eta_0^{-1} T_{\mu\nu} (A_\alpha) =
-{1\over 2} g_{\mu\nu} A_{\alpha;\beta}A^{\alpha;\beta} +
 A_{\mu ;\alpha}A_\nu^{;\alpha} + A_{\alpha;\mu}A^\alpha_{;\nu} -
\nonumber \\
{1\over 2}\left( A_{\mu;\alpha} A_\nu +A_{\nu;\alpha} A_\mu +
A_{\alpha;\mu} A_\nu + A_{\alpha;\nu} A_\mu -
A_{\mu;\nu} A_\alpha - A_{\nu;\mu} A_\alpha \right)^{;\alpha}
\label{taa}
\end{eqnarray}
The Hubble parameter which enters equation (\ref{d2tat}) is determined by the
expression:
\beq{
3H^2 m_1^2  = \rho_{tot} = \rho_{vac} + T_{tt}
\label{hat}
}\eeq
where $m_1^2 = m_{Pl}^2/8\pi$.

We will consider a special homogeneous solution: $A_j =0$ and $A_t =A(t)$. We
assume that initially the magnitude of $A_t$ is small and the expansion of the
universe is dominated by the vacuum energy,
$H_v = \sqrt{ 8\pi \rho_{vac} /3m_{Pl}^2}$. In this regime $A_t$ exponentially
rises, $A_t(t) \sim \exp (0.79 Ht)$ and soon its contribution into the energy
density becomes non-negligible. If $\eta_0 = - 1$ is chosen so that
the vacuum energy
density and the energy density of the field $A_t$ has opposite signs, the
contribution of $A_t$ would diminish $H$ and both the expansion rate
and the rate
of increase of $A_t$ would slow down. One can check that asymptotically
$A_t \sim t$ and $H=1/t$. Expanding the solution in powers of $1/t$ and
assuming that $\rho_{vac}> 0$ and $\eta_0 = -1<0$  we find:
\beq{
A_t = t\sqrt{\rho_{vac} /2} \left( 1 + {c_1 \over t} + {c_2 \over t^2}
\right)
\label{aasym}
}\eeq
\beq{
H = {1\over t} \left( 1 - {c_1 \over t} + {c_1^2 - 4c_2/3 \over t^2}\right)
\label{hasym}
}\eeq
where $c_2 = 3m_{Pl}^2/8\pi \rho_{vac}$ and $c_1$ is determined by initial
conditions. The energy and pressure density of this solution are respectively
\beq{
\rho(A_t) = {1\over 2}\dot A_t^2 + {3\over 2} H^2A^2_t \rightarrow
\rho_{vac} (-1 + c_2/t^2)
\label{rhoat}
}\eeq
and
\beq{
p(A_t) = \rho_{vac} (1 - c_2/3t^2)
\label{pat}
}\eeq

>From eq.(\ref{rhoat}) we obtain the following expression for the Hubble
parameter:
\beq{
H^2 = {\rho_{vac} + \eta_0 \dot A_t^2/2 + \rho_{matter}
\over 3(m^2_1 -\eta_0 A_t^2/2)}
\label{h2}
}\eeq
The energy density of normal matter, $\rho_{matter}$, is added
here for generality. Since
$\rho_{matter} \sim 1/a^4$ for relativistic matter, $\rho_r$,
and $\sim 1/a^3$ for non-relativistic matter, $\rho_{nr}$, the contribution
of the usual matter into total cosmological energy density quickly dies down,
$\rho_r\sim 1/t^4$ and $\rho_{nr}\sim 1/t^3$, and becomes negligible. Thus the
result $H\ = 1/t$ does not depend on the matter content and follows from the
asymptotic rise of the field, $A_t \sim t$. The total
cosmological energy density in this model is dominated by the remnant of
$(\rho_{vac}-\rho_A )\sim 1/t^2$. This cosmology is not realistic and
this is because
the expansion rate, $a(t) \sim t$ is too fast. One can try to construct a model
with a slower expansion rate using the freedom of adding new derivative terms
into the Lagrangian:
\beq{
{\cal L}_1 =\eta_1 A_{\mu;\nu}A^{\nu;\mu}
\label{l1}
}\eeq
\beq{
{\cal L}_2 =\eta_2 (A^\mu\>_{;\mu})^2
\label{l2}
}\eeq
However the first one gives exactly the same equation of motion for $A_t$ as
the Lagrangian ${\cal L}_0$ and the contribution from ${\cal L}_2$ into
the equation of motion is just $ \eta_2 A^\alpha_{\>;\alpha;\mu}$. It does not
change the asymptotic behavior obtained above.
So for a more realistic cosmologies one
has to address to higher rank fields. We will do that in the next section.

Let us consider now the contribution of the space components $A_j$ into the
energy density. It follows from eq. (\ref{dt2aj}) that in the cosmological
background with $H=1/t$ the space components
$A_j$ increase as $t^{\sqrt 2}$ i.e. even faster
than $A_t$, but the energy density of these components remain small in
comparison with $\rho (A_t) \approx const$ (\ref{rhoat}):
\beq{
\rho (A_j ) =  {1\over a^2} \left( -{1\over 2} \dot A_j^2 + H \dot A_j A_j
-H^2 A_j^2 \right) \sim t^{2\sqrt{2} -4} = t^{-1.17}
\label{rhoaj}
}\eeq
However since $\rho (A_t)$ is canceled with $\rho_{vac}$ up to terms of the
order $1/t^2$ the contribution of $\rho (A_j)$ becomes dominant. Moreover
the energy-momentum tensor of $A_j$ contains undesirable non-isotropic terms
proportional to $A_iA_j$ or to $\dot A_i A_j$.
These terms can be suppressed if one adds the
Lagrangian ${\cal L}_1$ (\ref{l1}) with the proper choice of parameter
$\eta_1$.
One can check that in this case the space components rise as
$A_j \sim t^{\sqrt{ 2(1+\eta_1/\eta_0)}}$. So for $-1<\eta_1/\eta_0 <-1/2$ the
contribution of $A_j$ into cosmological energy density would be small. Though
the model of this Section is not realistic the tricks used here may be useful
for more realistic models considered in the following section.

One more comment about the cosmological solutions with $A_t$ may be of
interest.
Let us assume now that $\eta_0$ is positive, $\eta_0 = 1 $. Corresponding
cosmological model in this case possesses a rather peculiar singularity.
The equation of motion
(\ref{d2tat}) does not change and the field $A_t$ remains unstable in the
Robertson-Walker background but the behavior of the solution becomes quite
different. One can see from eq.(\ref{h2}) that the Hubble parameter $H$ has
a singularity during expansion stage at a finite value of the field amplitude
and at a finite time. The solution near the singularity has the form:
\beq{
H = {h_1 \over (t_0 - t)^{2/3} },
\label{hsin}
}\eeq
\beq{
 A_t(t) =  \sqrt 2 m_1 \left[1 + c_1 \left( t_0 - t\right)^{2/3} \right]
\label{atsin}
}\eeq
where $m_1 = m_{Pl} /\sqrt{8\pi}$ and $c_1$ and $h_1$ are constant.
The energy density of the field $A_t$ at the singular point
tends to infinity as $(t_0 - t)^{-2/3}$
while the scale factor tends to constant value according to the expression
$a(t) \sim \exp [-3h_1 (t_0 -t )^ {1/3}]$. Since this is not related to the
problem of the cosmological constant we will not go into further details and
postpone the discussion of the solution with positive $\eta_0$ for the future.

\section{Higher rank symmetric fields.}

Essential features of cosmologies with higher rank symmetric tensor fields are
the same as discussed in the previous section
but some details may be different and in particular the expansion
rate. Equation of motion (\ref{d2v}) for the space-point independent
components of the second
rank symmetric tensor $S_{\alpha\beta} $ in the flat RW background (\ref{ds2})
has the form:
\beq{
 (\partial_t^2 + 3H\partial_t -6H^2) S_{tt} -2H^2 s_{jj} = 0
\label{dt2tt}
}\eeq
\beq{
 (\partial_t^2 + 3H\partial_t -6H^2) s_{tj} = 0
\label{dt2tj}
}\eeq
\beq{
 (\partial_t^2 + 3H\partial_t -2H^2) s_{ij} -2H^2 \delta_{ij} S_{tt} = 0
\label{dt2ij}
}\eeq
where  $s_{tj} = S_{tj}/a(t)$ and $s_{ij} = S_{ij}/a^2(t)$.

For $\eta_0 = -1$ there exists a particularly interesting homogeneous
solution of these
equations which at large $t$ behaves as $S_{tt} = Ct$, $s_{ij} =
\delta_{ij} Ct /3$, and $s_{tj} =0$. The condition of vanishing of $s_{tj}$
is not stable and we will return to that below. There may be nonvanishing
components $s_{ij}$ which are not proportional to the isotropic tensor
$\delta_{ij}$ but they rise with time slower than $t$.
The energy density corresponding to this solution
\beq{
 \rho = \eta_0\left[{1\over 2} (\dot S^2_{tt} + \dot s^2_{ij} )
+ H^2 (3S^2_{tt} + s_{ij}^2 + 2 S_{tt}s_{jj} )\right]
\label{rhostt}
}\eeq
exactly compensates the vacuum energy density, as above in the case of vector
field, but the expansion rate at large $t$ is different:
\beq{
 H = {3\over 8t}
\label{3h8t}
}\eeq
In this model $a\sim t^{3/8}$ and the energy density of usual matter decreases
rather slowly, $\rho_r \sim t^{-3/2}$ and $\rho_{nr} \sim t^{-9/8}$.
Corresponding values of the parameter $\Omega = \rho_{matter} /\rho_c$ would
be much larger than 1. Though the energy density of the usual matter may be
the dominant one, the Hubble parameter, as above,
 does not depend on it. Using expression
(\ref{rhostt}) we find similar to (\ref{h2}):
\beq{
 H^2 = { \rho_{vac} + \eta_0 (\dot S_{tt}^2 + \dot s_{ij}^2) /2 +\rho_{matter}
 \over 3 m_1^2 - \eta_0 (3S_{tt}^2 + s^2_{ij} + 2S_{tt}s_{jj} ) }
\label{h22}
}\eeq
One can easily check that the asymptotic solution of the equation of motion
$S_{tt}\sim t$ and $s_{ij} \sim t$ gives the result (\ref{3h8t}) independently
of the matter content and its properties, if only $\rho_{matter} $ decreases in
the course of expansion. This is in a drastic contrast to the standard
cosmology, when the expansion rate is determined by the usual matter, so for
an agreement with observations a particular fine-tuning is necessary even if
one manages to obtain a normal expansion rate. To achieve the latter we can use
the freedom in the choice of the Lagrangian of the tensor field similar to
expressions (\ref{l1}) and (\ref{l2}):
\beq{
\Delta {\cal L} = \eta_1 S_{\alpha\beta;\gamma} S^{\alpha\gamma;\beta}
+\eta_2 S^\alpha_{\beta;\alpha} S^{\gamma\beta}_{\> \>;\gamma}
+\eta_3 S^\alpha_{\alpha;\beta} S_\gamma^{\gamma;\beta}
\label{deltal}
}\eeq

The corresponding equations of motion can be written as:
\begin{eqnarray}
(\partial^2_t + 3H\partial_t - 6H^2)S_{tt} - 2H^2 s_{jj} +
\nonumber \\
 C_1 [(\partial^2_t + 3H\partial_t - 3H^2)S_{tt} + (H\partial_t - H^2) s_{jj}]
+
\nonumber \\
C_2 \partial_t [(\partial_t + 3H) S_{tt} + Hs_{jj}] +
\nonumber \\
C_3 (\partial^2_t + 3H\partial_t ) (S_{tt} - s_{jj}) = 0,
\label{stt}
\end{eqnarray}
 \begin{eqnarray}
(\partial^2_t + 3H\partial_t - 6H^2)s_{tj} +
\nonumber \\
( C_1/2 ) (\partial^2_t + 3H\partial_t  -2 \dot H - 12H^2)s_{tj} +
\nonumber \\
(C_2/2 ) [\partial_t (\partial_t + 4H) s_{tj} - H(\partial_t + 4H) s_{tj}]= 0,
 \label{stj}
\end{eqnarray}
 \begin{eqnarray}
(\partial^2_t + 3H\partial_t - 2H^2)s_{ij} - 2H^2\delta_{ij} S_{tt} -
\nonumber \\
 C_1 [( \dot H + 4 H^2)s_{ij} + \delta_{ij} (H\partial_t +\dot H +
 4 H^2) S_{tt} ] -
\nonumber \\
C_2 \delta_{ij} [ H^2 s_{ll} + ( H\partial_t + 3H^2) S_{tt}] -
\nonumber \\
 C_3 (\partial^2_t + 3H\partial_t ) (S_{tt} - s_{jj}) = 0,
 \label{sij}
\end{eqnarray}
where $C_j =\eta_j /\eta_0$.

As before these equations have unstable solutions $S_{tt}\sim t$ and
$s_{ij}\sim \delta_{ij}t$ with $H \sim 1/t$; this solution annihilates
the vacuum energy. Varying the parameters $\eta_j$
we may obtain different expansion regimes and in particular
the relativistic expansion $H=1/2t$ or the non-relativistic one, $H=2/3t$.
An interesting choice is $C_1 + C_2 = -1$ because it ensures $s_{tj}=0$. The
Hubble parameter corresponding to this choice is given by
\beq{
 Ht = { 3(C_1 +1) + C_3 (15 +16 C_1) \over  3(C_1 +1) + C_3 (20 + 16 C_1)}
\label{ht}
}\eeq
Taking e.g. $C_1 = 0 $ and $C_3 = -3/10$ we get $H= 1/2t$ as in the
radiation dominated universe. So
the cosmology does not look as unrealistic as the one in the preceding section,
though the reason for the choice of the particular values of $C_j$ is
absolutely unknown.
In this cosmological model the energy density of relativistic matter and the
non-compensated part of the energy density of the $S_{\alpha\beta}$,
$\delta \rho = \rho_{vac}-\rho(S_{\alpha\beta})$ both decrease with time as
$1/t^2$ and if initially (at $t=t_{Pl} = 1/m_{Pl}$) $\rho_r \sim m^4_{Pl} $
and if $\rho_{vac} $ has similar magnitude, then the contribution of
relativistic matter into cosmological parameter $\Omega$:
$\Omega_r = \rho_r /(3H^2m_{Pl}^2 /8\pi)$ would be always around unity. Since
it is rather natural to assume that all initial values were close to the Planck
ones the equality of $\rho_r$ and $\delta \rho$ does not imply too much of
fine-tuning. However the successful results of the theory of primordial
nucleosynthesis demands a rather restrictive relation between the rate of
the expansion and the energy density of relativistic matter. In this model
it is a free parameter which should be rather precisely chosen. This issue is
related to the problem of matter creation in the early universe and will be
addressed elsewhere. (For a recent analysis of the nucleosynthesis bounds
on cosmologies with varying cosmological "constant" see ref.\cite{sar}.)

Another and a more serious problem of this model is that the Hubble parameter
does not depend upon the matter content of the universe, as it has
been mentioned above. Correspondingly the expansion would always remain
relativistic and this may create some rather evident cosmological problems.
It may be interesting to construct a model in which the parameters $\eta_j$
vary together with the expansion in such a way that the regime changes from
the relativistic to nonrelativistic one (for the case considered above it
is achieved for $C_3 = -3/5$). However a natural mechanism for
realization of such a scenario is not found. Still though the model does not
look realistic one may hope that in a more complicated version it will describe
our universe.

\section{Discussion.}

There are two questions which are vitally important for the proposed model.
First one is rather philosophical, whether it worthwhile to make such a
construction for killing vacuum energy. The assumptions made above about the
form of the Lagrangian are not absolutely harmless. Quantum version of this
model would definitely meet very strong difficulties like negative
probabilities, nonrenormalizability, etc. Quantum corrections may dramatically
change the form of the Lagrangian in particular generating nonzero masses
and/or some other terms which may depend on the field but not its derivative.
If so, the model would not work.

On the other hand there is absolutely no way, known at the present day, how
one can get rid of huge vacuum energy which would make our life impossible in
such a universe. One can of course invoke the antropic principle but in many
cases it is very close to the assumption of a supernatural creation of the
world. If this is the case cosmologists would be jobless.
This model at least propose a mechanism of automatic
cancellation of $\rho_{vac}$ with time independent gravitational constant,
which was a drawback of previous attempts. It gives reasonable
theoretical frameworks
(at classical level) for cosmology with asymptotically vanishing vacuum
energy based on Lagrangian approach.

The second question is if it is possible at all to construct a realistic
cosmology based on this approach. There is a subquestion if the realistic
model would be natural or demands a strong fine-tuning like the fine-tuning of
the vacuum energy in the traditional cosmology. The answer to the last part
of the question is semi-negative. At the moment no way is seen how to make a
natural model but the necessary fine-tuning is possibly not as strong as
$10^{100}$. On the other hand one may impose ad hoc the necessary values of
the parameters and to use this model as a toy one to study  cosmology with
automatic cancellation of $\Lambda$. In particular the model considered here
gives the non-compensated amount of vacuum energy of the order of
$\rho_c \sim m^2_{Pl}/t^2$. As it was argued in ref.\cite{ad2,ad3}
this is a generic phenomenon. One more problem which may be inherent
for this kind of cosmological
model is that inflation is very difficult or even impossible to realize. The
vacuum (or vacuum-like) energy is destroyed so fast that inflation stops before
it starts. To avoid it one has to suppress somehow the instability of the
fields when their amplitude is small. More work in this and many other
directions is necessary.

\bigskip
{\bf Acknowledgment.}
This paper was supported in part by the Danish National Science Research
Council
through grant 11-9640-1 and in part by Danmarks Grundforskningsfond through its
support of the Theoretical Astrophysical Center.

\end{document}